# Increasing the resolution of transmission electron microscopy by computational ghost imaging

P. Rosi[1,2], L. Viani[2], E. Rotunno[1]*, S. Frabboni[1,2], A. H. Tavabi[3], R. E. Dunin-Borkowski[3], A. Roncaglia[4], V. Grillo[1]

1) Institute of Nanosciences CNR-S3, via G.Campi 213, 41125 Modena, Italy
2) FIM Department, University of Modena and Reggio Emilia, via G. Campi 213/A, 41125 Modena, Italy
3) Ernst Ruska-Centre for Microscopy and Spectroscopy with Electrons, Forschungszentrum Jülich, 52425 Jülich, Germany
4) Institute for Microelectronics and Microsystems- CNR, Via P. Gobetti, 101, 40129 Bologna, Italy

**Abstract**

By means of numerical simulations, we demonstrate the innovative use of computational ghost imaging in transmission electron microscopy to retrieve images with a resolution that overcomes the limitations imposed by coherent aberrations. The method requires measuring the intensity on a single pixel detector with a series of structured illuminations. The success of the technique is improved if the probes are made to resemble the sample and the patterns cover the area of interest evenly. By using a simple 8 electrode device as a specific example, a 2-fold increase in resolution beyond the aberration limit is demonstrated to be possible under realistic experimental conditions.

**Main Text**

Transmission electron microscopes allow for imaging with high lateral resolution. Unlike light microscopes, the main resolution-limiting factor is not the probe wavelength but aberrations. Since the time of Scherzer [1], it has been known that a system of rotationally-symmetric lenses has unavoidable positive spherical aberration (Cs), which can be compensated by a complex system of additional lenses and multipoles [2–4]. These aberration correctors can now be used to correct aberrations up to 6th order and are considered to be mandatory, albeit expensive, tools. [5]

However, if the imaging system is coherent, then no information is lost due to aberrations and the retrieval of a high-resolution image of the sample is still possible by using computational procedures, such as holography [6,7], focal series reconstruction [8–10], the transport of intensity equation [11] and ptychography [12–14]. The resolution of ptychography does not depend on the probe size, which is usually defocused intentionally [15]. Instead, it results from the collection of diffraction from the sample over a large angular range, followed by retrieval of the phase using inverse methods [14,16,17]. The most spectacular success of ptychography is the achievement of 0.2 Å resolution, with the main limiting factor being atomic thermal vibration [18].

Techniques that are referred to as computational ghost imaging [19,20] (CGI) and single pixel imaging [21] follow a complementary approach, which has been explored thoroughly in light microscopy [22–30]. Without using a camera, spatial information is inferred through structuring of the probing beam and the massive use of computation. A single-pixel bucket detector integrates the transmitted/ reflected intensity signal generated by interaction of the sample with each unique pattern in a series of known different structured illuminations. CGI is then used to retrieve the sample transmission function computationally.

Each technique that makes use of correlation between the probe shape and the recorded intensity can be considered as a form of CGI, even conventional scanning transmission electron microscopy (STEM). STEM involves the use of a series of trivially-structured (point-like) illuminations and the collection of scalar signals using single pixel detectors, such as bright-field and annular dark-field (ADF) detectors. The added value of CGI is that it potentially encompasses not only the elastically scattered ADF signal, but also a variety of other single pixel signals, such as energy dispersive X-ray, electron energy-loss and cathodoluminescence signals. For each of these signals, CGI can be used to overcome the aberration-limited resolution by incorporating aberrations in the structuring of the probe and, consequently, in the reconstruction scheme.

Unfortunately, electron CGI is still at an early stage [31], largely due to the difficulty of tuning the electron wavefunction arbitrarily. Electron spatial modulators that can be used to control amplitude and phase freely have so far only been demonstrated for specific conditions and with limited scope [32–35].

In a recent paper [36], we demonstrated experimentally that a micro electromechanical systems (MEMS) based electron modulator can be used to provide, in its far field, a crude but effective electron modulation that can be used for CGI. Here, we propose a new design consisting of 4 pairs of electrodes, which are distributed radially 90° apart from each other. When such an electron phase modulator is illuminated by a plane wave in the condenser system of the TEM (Fig. 1a), it generates a probe in the sample plane that can be understood as the Fourier transform of the beam's wavefunction at the device plane.

We use numerical simulations to demonstrate a resolution that is comparable to Cs-corrected STEM imaging, by using the modulator to perform CGI on an uncorrected instrument (Fig. 1b-g). We simulated such a CGI experiment for an $MoS_2$ twisted bilayer (Fig. 1b), in which two $MoS_2$ monolayers are stacked on top of each other at a 7° angle. Such a sample offers an almost continuous range of interatomic distances and has proven to be useful to estimate the resolution of imaging techniques [37]. We assume a 300 kV microscope with $Cs = 2.7$mm which requires, under Scherzer conditions, the use of a small aperture ($\alpha = 7.3\ mrad$), limiting the resolution to 1.63 Å, as shown in Fig. 1d. In contrast, CGI allows for the use of a larger numerical aperture, thereby achieving higher resolution. In Fig. 1f-g we show a CGI reconstruction computed using a $2\alpha$ convergence semi-angle, providing a resolution of 0.64 Å despite the presence of aberrations. This value is comparable to the resolution of an aberration-free imaging system with the same semi-convergence angle. An arbitrarily large semi-convergence angle can in fact be used, provided the ability to predict the illumination patterns. An accurate probe prediction is essential for future experimental realization, as discussed in the Supplemental Material [38]. A limiting factor is coherence [38], whose damping envelope defines the highest frequency transmitted to the structured illumination [39,40].

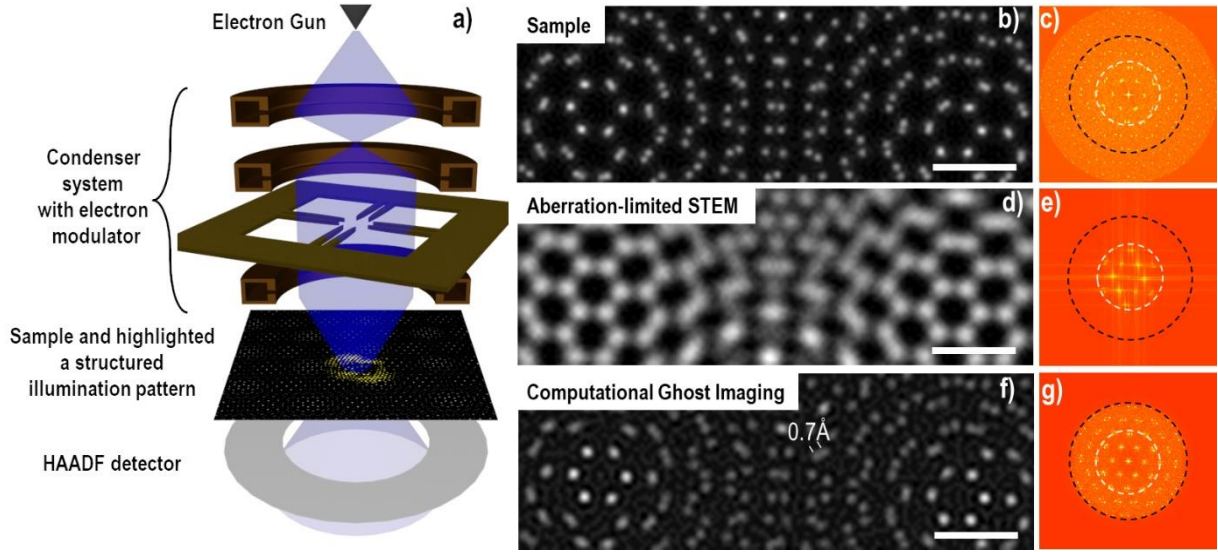

*Figure 1: a) Schematic representation of the electron-optical setup. b), d), f) Comparison between an MoS$_2$ twisted bilayer atomic potential, an aberration-limited STEM image, a CGI reconstruction and c), e), g) their FFTs, in which the light circle corresponds to a semi-convergence angle of $2\alpha$, while the dark circle corresponds to $4\alpha$. The scale bars are 5 Å.*

In its present form, the CGI imaging scheme assumes a linear measurement, *i.e.*, that the measured signal $I$ corresponds to the overlap integral (scalar product) between the probe and the sample transmission object function $\vec{T}$. In matrix notation, $\vec{I} = \hat{P}\vec{T}$, where $\vec{I}$ is an array containing the $N$ signals recorded by the single pixel detector, corresponding to $N$ different illuminations, which are gathered in $\hat{P}$, the array containing the intensities of the $N$ probes. Many algorithms exist to solve for $\vec{T}$ [41–43], some of which are compared in the Supplemental Material [38]. Here, the conjugate gradient descent method [42] is used.

Independently from the algorithm used, a high-quality reconstruction is obtained more easily if the following two conditions are met, in addition to that of linearity. First, the illumination patterns should resemble the target to produce higher measured intensities. Second, the illumination intensity at the same point across different patterns should be independently identically distributed (i.i.d). The relevance of the first condition has been demonstrated by the light optics community, who have explored the best set of illumination patterns for specific scenes or applications [24,44–48]. In the context of atomic resolution imaging, the similarity condition corresponds to illumination patterns that resemble an atomic lattice, *e.g.*, featuring high contrast intensity fringes. The fringe spacing has to be on the same length-scale as the interatomic distances and is conventionally dictated only by the beam semi-convergence. However, the electron modulator bias configuration allows for further tuning of the spacing. The measured signal can be used to quantify the similarity of the patterns with the sample: the higher it is, the more similar they are.

The electron modulator in Fig. 1a is designed to achieve this situation. Each pair of electrodes behaves as a spiral phase plate (SPP) [49–51], which is able to produce a vortex-like phase gradient that winds $\ell$ times, where $\ell$ is referred to as the topological charge. Interference of the vortices is responsible for the fringe patterns. In particular, the correct arrangement of topological charges [38] generates a centrosymmetric phase distribution (Fig. 2a), which causes the probe wavefunction to be real-valued [52,53]. The modulator's phase is then translated into a high frequency intensity oscillation in the probe (Fig. 2b). This behavior is important, since phase oscillations would be of no use, as CGI is inherently an amplitude-based technique.

The aberrations in the probe-forming lens system add up to the phase produced by the electron modulator, implying that the effect of the known Cs can be accounted for by including it in the phase of the electron modulator (Fig. 2c). The presence of Cs (as well of other aberrations) alters the shape of the patterns, but they are still characterized by well-defined intensity fringes (Fig. 2d) and can be used for CGI reconstruction purposes.

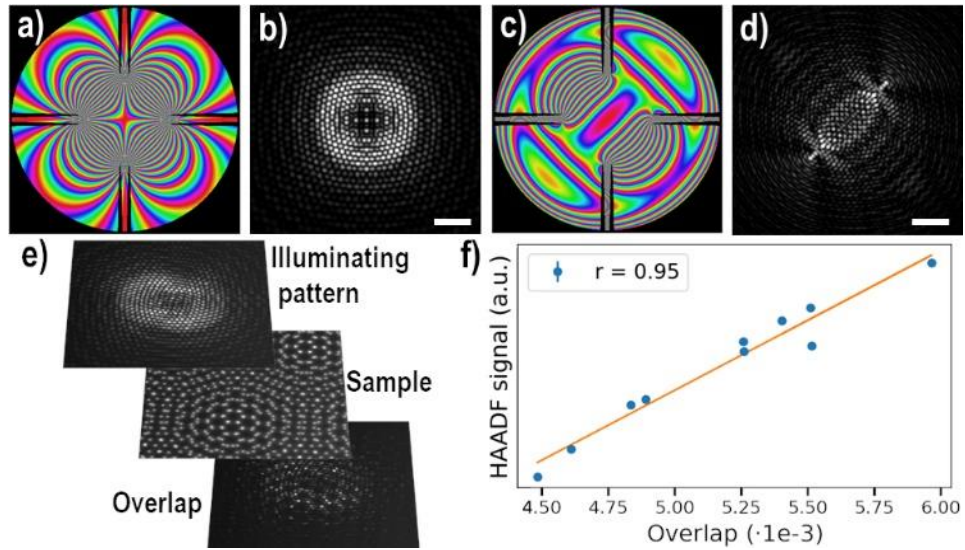

Figure 2: a) Phase profile produced by the electron modulator in a centrosymmetric condition without the effect of Cs and c) with Cs=2.7mm, and b), d) the corresponding probe shapes; e) Similarity between the patterns and the sample crystal structure, maximizing the overlap. f) ADF signal computed using a multislice simulation plotted as a function of the overlap of the probe with a 19.2-nm-thick Si [110] sample. In the legend is reported the Pearson correlation coefficient (r).

A CGI experiment requires a set of many patterns, whose number is proportional to the number of pixels in the image. Such a set was created by randomly changing the needles' bias in a prescribed range around the centrosymmetric condition. Small deviations from this condition produce deflections and stigmation of the beam, allowing for a larger area to be covered at the cost of the formation of caustics, which concentrate the intensity in a few points and may hamper the reconstruction.

We defined figures of merit, which are described in the Supplemental Material [38] and can be used to pre-select an optimal set of illumination patterns from the randomly-generated set that satisfy the conditions of i.i.d and similarity to the sample. All of the examples of structured illumination patterns shown in the main text and the Supplemental Material were obtained by satisfying these conditions. We built a database comprising 200 000 structured illumination patterns, sampled on a 400 x 400 point grid, which uniformly covers an area of 10 x 10 nm [38].

In this work, we assume an ADF (in particular high-angle) detector as our single pixel detector, and that the single pixel signal is computed as a superposition integral between the probe intensity and the squared lattice potential (Figure 2e), which is an explicit consequence of the incoherent nature of the ADF signal and of the fact that we neglect, for a sample that is thin or nearly homogeneous, any strong dynamical effects of the probe [54]. The possible breakdown of these approximations also affects the recorded intensity in standard STEM-ADF imaging. We further confirmed these approximations and the linearity of the measurement from the correlation between the assumed single pixel signal and the ADF signal, which we simulated using a multislice algorithm [55,56]. (Figure 2f shows the correlation of the two signals for a 55 mrad inner detection semi-angle and a 20-nm-thick Si [110] benchmark sample. Here, good agreement is shown between the intensity predicted by the multislice algorithm and a simple linear approximation. For such high scattering angles and low values of sample thickness, all inelastic scattering apart from phonon scattering can be neglected [57]. Inelastic scattering originating from plasmon losses does not, in fact, contribute strongly to the signal, but mainly to a small absorption effect that does not affect the GI reconstruction significantly.

Even though linearity is verified up to a thickness of ~20 nm, in this paper we limit ourselves to a bilayer sample, for which the linearity approximation is satisfied more easily. The signal is simulated for different beam currents in the range 5-1000 pA and a dwell time per pattern of 100 µs. We also assume that the ADF detector has a quantum efficiency of 1, which is not an extreme approximation for modern ADF detectors, leaving counting noise as the only source of noise in the system.

Fig. 3a shows reconstructed sample images for different sampling ratios (SRs), i.e., the ratio between the number of measurements to the number of pixels in the image, for three beam currents.

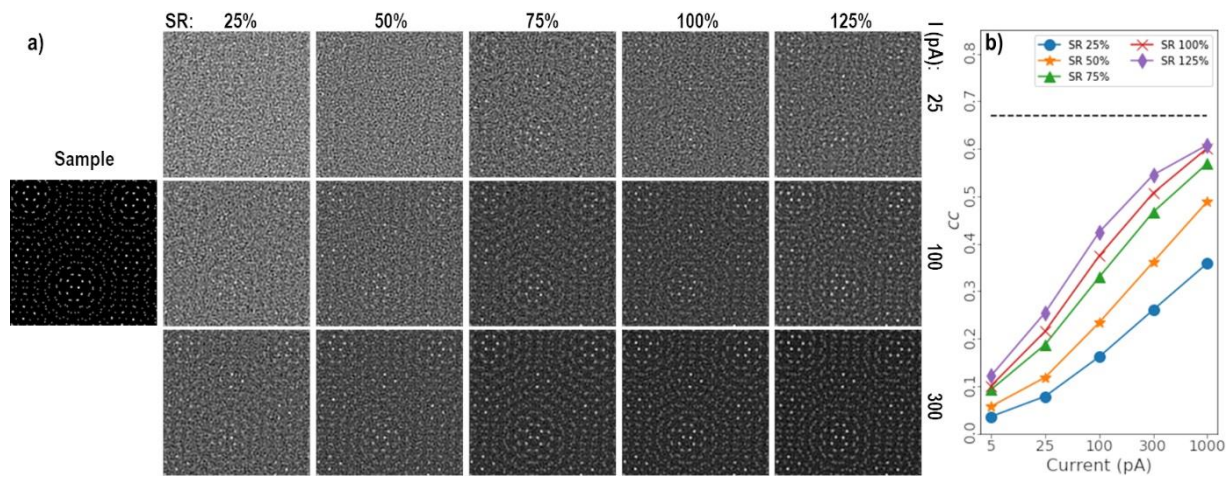

Figure 3: Effect of sampling ratio on reconstructed image quality for three electron currents. In a), the same region of a $3.6\ nm \times 3.6\ nm$ simulated sample is reconstructed. b) shows the behaviour of $cc$ as a function of beam current for different SRs. The black dashed line shows the maximum value of $cc$ obtained in the limit of infinite current.

In order to quantify the reconstruction accuracy, we make use of the cross-correlation coefficient defined by Thust and Urban [58] averaged over the entire image ($cc$), as reported in Fig.3b. The $cc$ values range between 0 and 1, where $cc = 0$ refers to zero correlation between the reconstructed image and the reference image and $cc = 1$ refers to a perfect match.

In general, CGI suffers in the low current regime from the well-known sensitivity of CGI to noise [59,60]. Increasing both the SR and the beam current improves the reconstruction quality. Interestingly, it is possible to achieve the same reconstruction quality by using either a high SR and a low current or a low SR and a high current, as shown in the form of a comparison between 25 pA at 125% SR in the top right corner and 300 pA at 25% SR in the lower left corner, both of which lead to high quality images. Even though these currents are higher than those used in typical high-resolution STEM experiments [61], the total electron dose can be contained since the reconstruction can be achieved with low SRs.

In conclusion, our numerical simulations show that CGI can be used to overcome the resolution limit imposed by the aberrations of the imaging system in a transmission electron microscope, thereby potentially eliminating the need to use an expensive aberration corrector. We propose the use of a MEMS-based electron modulator to produce illumination patterns that resemble a lattice structure. When used with an ADF single pixel detector, this scheme allows for the quantitative imaging of thin crystalline samples with a resolution that is comparable to ptychographic reconstruction.

We acknowledge support from the European Union's Horizon 2020 Research and Innovation Programme (grant agreement no. 964591 "SMART-electron") and HORIZON EUROPE framework program for research and innovation (grant agreement n. 101094299 "IMPRESS"). We further acknowledge support from the Italian MIR through the PRIN Project no. 2022249HSF "AI-TEM". The data and custom API that support the findings of this study are openly available in Zenodo at https://doi.org/10.5281/zenodo.8366329

**References**

[1] O. Scherzer, *Über Einige Fehler von Elektronenlinsen*, Zeitschrift Für Phys. **101**, 593 (1936).
[2] M. Haider, S. Uhlemann, E. Schwan, H. Rose, B. Kabius, and K. Urban, *Electron Microscopy Image Enhanced*, Nature **392**, 768 (1998).
[3] O. L. Krivanek, N. Dellby, and A. R. Lupini, *Towards Sub-Å Electron Beams*, Ultramicroscopy **78**, 1 (1999).
[4] H. H. Rose, *Historical Aspects of Aberration Correction*, J. Electron Microsc. (Tokyo). **58**, 77 (2009).
[5] P. W. Hawkes and O. L. Krivanek, *Aberration Correctors, Monochromators, Spectrometers*, in (2019), pp. 625–675.
[6] E. Völkl, L. F. Allard, and D. C. Joy, *Introduction to Electron Holography* (Springer Science & Business Media, 1999).
[7] A. Tonomura, *Electron Holography*, in (1999), pp. 29–49.
[8] R. R. Meyer, A. I. Kirkland, and W. O. Saxton, *A New Method for the Determination of the Wave Aberration Function for High Resolution TEM*, Ultramicroscopy **92**, 89 (2002).
[9] A. I. Kirkland and R. R. Meyer, *"Indirect" High-Resolution Transmission Electron Microscopy: Aberration Measurement and Wavefunction Reconstruction*, Microsc. Microanal. **10**, 401 (2004).
[10] R. R. Meyer, A. I. Kirkland, and W. O. Saxton, *A New Method for the Determination of the Wave Aberration Function for High-Resolution TEM.*, Ultramicroscopy **99**, 115 (2004).
[11] M. R. Teague, *Deterministic Phase Retrieval: A Green's Function Solution*, J. Opt. Soc. Am. **73**, 1434 (1983).
[12] P. Thibault, M. Dierolf, A. Menzel, O. Bunk, C. David, and F. Pfeiffer, *High-Resolution Scanning X-Ray Diffraction Microscopy*, Science (80-. ). **321**, 379 (2008).
[13] P. Thibault, M. Dierolf, O. Bunk, A. Menzel, and F. Pfeiffer, *Probe Retrieval in Ptychographic Coherent Diffractive Imaging*, Ultramicroscopy **109**, 338 (2009).
[14] A. M. Maiden and J. M. Rodenburg, *An Improved Ptychographical Phase Retrieval Algorithm for Diffractive Imaging*, Ultramicroscopy **109**, 1256 (2009).
[15] O. Bunk, M. Dierolf, S. Kynde, I. Johnson, O. Marti, and F. Pfeiffer, *Influence of the Overlap Parameter on the Convergence of the Ptychographical Iterative Engine*, Ultramicroscopy **108**, 481 (2008).
[16] R. W. Gerchberg and W. O. Saxton, *A Practical Algorithm for the Determination of Phase from Image and Diffraction Plane Pictures*, Optik (Stuttg). **35**, 237 (1971).
[17] J. M. Rodenburg and R. H. T. Bates, *The Theory of Super-Resolution Electron Microscopy via Wigner-Distribution Deconvolution*, Philos. Trans. R. Soc. London. Ser. A Phys. Eng. Sci. **339**, 521 (1992).
[18] Z. Chen, Y. Jiang, Y.-T. Shao, M. E. Holtz, M. Odstrčil, M. Guizar-Sicairos, I. Hanke, S. Ganschow, D. G. Schlom, and D. A. Muller, *Electron Ptychography Achieves Atomic-Resolution Limits Set by Lattice Vibrations*, Science (80-. ). **372**, 826 (2021).
[19] J. H. Shapiro, *Computational Ghost Imaging*, Phys. Rev. A **78**, 061802 (2008).
[20] Y. Bromberg, O. Katz, and Y. Silberberg, *Ghost Imaging with a Single Detector*, Phys. Rev. A **79**, 053840 (2009).
[21] M. F. Duarte, M. A. Davenport, D. Takhar, J. N. Laska, T. Sun, K. F. Kelly, and R. G. Baraniuk, *Single-Pixel Imaging via Compressive Sampling*, IEEE Signal Process. Mag. **25**, 83 (2008).
[22] G. Futia, P. Schlup, D. G. Winters, and R. A. Bartels, *Spatially-Chirped Modulation Imaging of Absorbtion and Fluorescent Objects on Single-Element Optical Detector*, Opt. Express **19**, 1626 (2011).
[23] G. A. Howland, P. B. Dixon, and J. C. Howell, *Photon-Counting Compressive Sensing Laser Radar for 3D Imaging*, Appl. Opt. **50**, 5917 (2011).

[24] R. I. Stantchev, B. Sun, S. M. Hornett, P. A. Hobson, G. M. Gibson, M. J. Padgett, and E. Hendry, *Noninvasive, near-Field Terahertz Imaging of Hidden Objects Using a Single-Pixel Detector*, Sci. Adv. **2**, (2016).

[25] R. I. Stantchev, J. C. Mansfield, R. S. Edginton, P. Hobson, F. Palombo, and E. Hendry, *Subwavelength Hyperspectral THz Studies of Articular Cartilage*, Sci. Rep. **8**, 6924 (2018).

[26] M.-J. Sun and J.-M. Zhang, *Single-Pixel Imaging and Its Application in Three-Dimensional Reconstruction: A Brief Review*, Sensors **19**, 732 (2019).

[27] J. S. Totero Gongora, L. Olivieri, L. Peters, J. Tunesi, V. Cecconi, A. Cutrona, R. Tucker, V. Kumar, A. Pasquazi, and M. Peccianti, *Route to Intelligent Imaging Reconstruction via Terahertz Nonlinear Ghost Imaging*, Micromachines **11**, 521 (2020).

[28] L. Olivieri, J. S. T. Gongora, L. Peters, V. Cecconi, A. Cutrona, J. Tunesi, R. Tucker, A. Pasquazi, and M. Peccianti, *Hyperspectral Terahertz Microscopy via Nonlinear Ghost Imaging*, Optica **7**, 186 (2020).

[29] S.-C. Chen, Z. Feng, J. Li, W. Tan, L.-H. Du, J. Cai, Y. Ma, K. He, H. Ding, Z.-H. Zhai, Z.-R. Li, C.-W. Qiu, X.-C. Zhang, and L.-G. Zhu, *Ghost Spintronic THz-Emitter-Array Microscope*, Light Sci. Appl. **9**, 99 (2020).

[30] L. Olivieri, L. Peters, V. Cecconi, A. Cutrona, M. Rowley, J. S. Totero Gongora, A. Pasquazi, and M. Peccianti, *Terahertz Nonlinear Ghost Imaging via Plane Decomposition: Toward Near-Field Micro-Volumetry*, ACS Photonics **10**, 1726 (2023).

[31] S. Li, F. Cropp, K. Kabra, T. J. Lane, G. Wetzstein, P. Musumeci, and D. Ratner, *Electron Ghost Imaging*, Phys. Rev. Lett. **121**, 114801 (2018).

[32] J. Verbeeck, A. Béché, K. Müller-Caspary, G. Guzzinati, M. A. Luong, and M. Den Hertog, *Demonstration of a 2 × 2 Programmable Phase Plate for Electrons*, Ultramicroscopy **190**, 58 (2018).

[33] P. Thakkar, V. A. Guzenko, P.-H. Lu, R. E. Dunin-Borkowski, J. P. Abrahams, and S. Tsujino, *Fabrication of Low Aspect Ratio Three-Element Boersch Phase Shifters for Voltage-Controlled Three Electron Beam Interference*, J. Appl. Phys. **128**, 134502 (2020).

[34] A. H. Tavabi, P. Rosi, E. Rotunno, A. Roncaglia, L. Belsito, S. Frabboni, G. Pozzi, G. C. Gazzadi, P.-H. Lu, R. Nijland, M. Ghosh, P. Tiemeijer, E. Karimi, R. E. Dunin-Borkowski, and V. Grillo, *Experimental Demonstration of an Electrostatic Orbital Angular Momentum Sorter for Electron Beams*, Phys. Rev. Lett. **126**, 094802 (2021).

[35] A. Konečná, E. Rotunno, V. Grillo, F. J. García de Abajo, and G. M. Vanacore, *Single-Pixel Imaging in Space and Time with Optically Modulated Free Electrons*, ACS Photonics **10**, 1463 (2023).

[36] A. Kallepalli, L. Viani, D. Stellinga, E. Rotunno, R. Bowman, G. M. Gibson, M.-J. Sun, P. Rosi, S. Frabboni, R. Balboni, A. Migliori, V. Grillo, and M. J. Padgett, *Challenging Point Scanning across Electron Microscopy and Optical Imaging Using Computational Imaging*, Intell. Comput. **2022**, (2022).

[37] Y. Jiang, Z. Chen, Y. Han, P. Deb, H. Gao, S. Xie, P. Purohit, M. W. Tate, J. Park, S. M. Gruner, V. Elser, and D. A. Muller, *Electron Ptychography of 2D Materials to Deep Sub-Ångström Resolution*, Nature **559**, 343 (2018).

[38] *See Supplemental Material at [URL Will Be Inserted by Publisher] for a Description of the Imaging Conditions Optimisation, Illumination Patterns Database, Reconstruction Algorithms and Application Limits.*

[39] P. D. Nellist and S. J. Pennycook, *Subangstrom Resolution by Underfocused Incoherent Transmission Electron Microscopy*, Phys. Rev. Lett. **81**, 4156 (1998).

[40] C. Dwyer, R. Erni, and J. Etheridge, *Measurement of Effective Source Distribution and Its Importance for Quantitative Interpretation of STEM Images*, Ultramicroscopy **110**, 952 (2010).

[41] D. G. Luenberger, *Introduction to Linear & Nonlinear Programming* (1973).

[42] M. R. Hestenes and S. Eduard, *Methods of Conjugate Gradients for Solving Linear Systems*, J. Res. Natl. Bur. Stand. (1934). **49**, 409 (1952).

[43] K. Guo, S. Jiang, and G. Zheng, *Multilayer Fluorescence Imaging on a Single-Pixel Detector*, Biomed. Opt. Express **7**, 2425 (2016).

[44] Z. Zhang, X. Wang, G. Zheng, and J. Zhong, *Hadamard Single-Pixel Imaging versus Fourier Single-Pixel Imaging*, Opt. Express **25**, 19619 (2017).

[45] G. M. Gibson, S. D. Johnson, and M. J. Padgett, *Single-Pixel Imaging 12 Years on: A Review*, Opt. Express **28**, 28190 (2020).

[46] S. Rizvi, J. Cao, K. Zhang, and Q. Hao, *DeepGhost: Real-Time Computational Ghost Imaging via Deep Learning*, Sci. Rep. **10**, 11400 (2020).

[47] X. Yu, R. I. Stantchev, F. Yang, and E. Pickwell-MacPherson, *Super Sub-Nyquist Single-Pixel Imaging by Total Variation Ascending Ordering of the Hadamard Basis*, Sci. Rep. **10**, 9338 (2020).

[48] L. López-García, W. Cruz-Santos, A. García-Arellano, P. Filio-Aguilar, J. A. Cisneros-Martínez, and R. Ramos-García, *Efficient Ordering of the Hadamard Basis for Single Pixel Imaging*, Opt. Express **30**, 13714 (2022).

[49] G. Pozzi, P.-H. Lu, A. H. Tavabi, M. Duchamp, and R. E. Dunin-Borkowski, *Generation of Electron Vortex Beams Using Line Charges via the Electrostatic Aharonov-Bohm Effect*, Ultramicroscopy **181**, 191 (2017).

[50] A. H. Tavabi, H. Larocque, P.-H. Lu, M. Duchamp, V. Grillo, E. Karimi, R. E. Dunin-Borkowski, and G. Pozzi, *Generation of Electron Vortices Using Nonexact Electric Fields*, Phys. Rev. Res. **2**, 013185 (2020).

[51] A. H. Tavabi, P. Rosi, A. Roncaglia, E. Rotunno, M. Beleggia, P.-H. Lu, L. Belsito, G. Pozzi, S. Frabboni, P. Tiemeijer, R. E. Dunin-Borkowski, and V. Grillo, *Generation of Electron Vortex Beams with over 1000 Orbital Angular Momentum Quanta Using a Tunable Electrostatic Spiral Phase Plate*, Appl. Phys. Lett. **121**, 073506 (2022).

[52] J. Guo, B. Guo, R. Fan, W. Zhang, Y. Wang, L. Zhang, and P. Zhang, *Measuring Topological Charges of Laguerre–Gaussian Vortex Beams Using Two Improved Mach–Zehnder Interferometers*, Opt. Eng. **55**, 035104 (2016).

[53] K. Y. Y. Bliokh, I. P. P. Ivanov, G. Guzzinati, L. Clark, R. Van Boxem, A. Béché, R. Juchtmans, M. A. A. Alonso, P. Schattschneider, F. Nori, and J. Verbeeck, *Theory and Applications of Free-Electron Vortex States*, Phys. Rep. **690**, 1 (2017).

[54] S. J. Pennycook, *Z-Contrast Stem for Materials Science*, Ultramicroscopy **30**, 58 (1989).

[55] J. M. Cowley and A. F. Moodie, *The Scattering of Electrons by Atoms and Crystals. I. A New Theoretical Approach*, Acta Crystallogr. **10**, 609 (1957).

[56] V. Grillo and E. Rotunno, *STEM_CELL: A Software Tool for Electron Microscopy: Part I—Simulations*, Ultramicroscopy **125**, 97 (2013).

[57] R. F. Egerton, *Electron Energy-Loss Spectroscopy in the Electron Microscope* (Springer Science & Business Media, 2011).

[58] A. Thust and K. Urban, *Quantitative High-Speed Matching of High-Resolution Electron Microscopy Images*, Ultramicroscopy **45**, 23 (1992).

[59] L. Bian, J. Suo, Q. Dai, and F. Chen, *Experimental Comparison of Single-Pixel Imaging Algorithms*, J. Opt. Soc. Am. A **35**, 78 (2018).

[60] W. Gong, *Performance Comparison of Computational Ghost Imaging versus Single-Pixel Camera in Light Disturbance Environment*, Opt. Laser Technol. **152**, 108140 (2022).

[61] R. Brescia, S. Toso, Q. Ramasse, L. Manna, J. Shamsi, C. Downing, A. Calzolari, and G. Bertoni, *Bandgap Determination from Individual Orthorhombic Thin Cesium Lead Bromide Nanosheets by Electron Energy-Loss Spectroscopy*, Nanoscale Horizons **5**, 1610 (2020).